\documentstyle[12pt]{article}
\begin{document}
\vspace{-0.2cm}
\begin{flushright}
Preprint IBRAE-95-08,\\
hep-th/9504139 \\
\vskip2mm
April 1995
\end{flushright}
\vskip2cm
\begin{center}
{ \bf SOME PROPERTIES OF THE KERR SOLUTION TO \\
\vskip3mm
LOW ENERGY STRING THEORY}\\
\vskip1cm
{\bf A.Ya. Burinskii} \footnote{ e--mail: grg@ibrae.msk.su}
\vskip2mm
{\em Gravity Research Group, Nuclear Safety Institute\\
Russian Academy of Sciences, B.Tulskaya 52, 113191 Moscow, Russia}
\end{center}
\vskip 2 cm
\centerline{\bf Abstract}
\par
\begin{quotation}
The Kerr solution to axidilaton gravity is  analyzed in the
Debney--Kerr--Schild formalism. It is shown that the Kerr principal null
congruence retains its property to be geodesic and shear free, however,
the axidilatonic Kerr solution is not algebraically special.
A limiting form of this solution is considered near the ring-like Kerr
singularity. This limiting solution coincides with the field of
fundamental heterotic string obtained by Sen \cite{sen,shet}.
\end{quotation}
\newpage
\section{Introduction}
Much attention has been paid recently to the connections of the black
hole physics and string theory. In particular, many of important solutions
of the Einstein gravity have  found their analogue among the
solutions of the low energy string theory including the axion and dilaton
corrections. Such classical solutions to the axidilaton gravity can be
interpreted as  stable extended soliton-like states  or fundamental strings
\cite{dab,sen,shet}.
In this paper we analyze a new rotating and charged solution to the
axidilaton gravity, which is an analogue of the Kerr solution.
This solution was obtained by Sen \cite{sen} and, in a more general form
(including the NUT parameter),
by Gal'tsov and Kechkin \cite{gal}. A rather complicated character
of the Kerr solution put obstacles in the way of direct obtaining
 this rotating solution from the field equations, so this solution was
obtained by a method for generating new solutions from the known
ones \cite{sha,sen,gal}. However, by using this method some
important characteristics of the new solutions remain unknown. For example,
there was no information concerning the type of the new Kerr-like solution
in the Petrov--Pirani classification.
\footnote{It was known only that this solution does not belong to the
type D in contrast to the Kerr solution of Einstein's gravity \cite{ggk}.}
We partially compensate for this deficiency.
\par
By using the Kerr coordinates \cite{dks} we analyze this solution
near the singular ring and find
the limiting form of the solution to coincide (up to some peculiarities)
with the solution constructed
by Sen \cite{sen,shet} for the field around a fundamental heterotic string.
\par
\section{Some algebraic properties of the Kerr solution to  axidilaton
gravity}
We will restrict in this paper to the case of electric charged Kerr
solution ( without the NUT-parameter and magnetic charge).
We are going to use the algorithm for obtaining this solution from the
original Kerr solution given by Gal'tsov and Kechkin \cite{gal}.
According to ref. \cite{gal}, starting with the vacuum Kerr solution
\begin{equation}
ds^2=\frac{\Delta-a^2 \sin^2\theta}{\Sigma}\left(dt-\omega
d\varphi\right)^2-\Sigma\left(\frac{dr^2}{\Delta}+d\theta^2+
\frac{\Delta\sin^2\theta}{\Delta-a^2\sin^2\theta} d\varphi^2\right),
\end{equation}
where
\begin{equation}
\Delta=r(r-2M) + a^2 ,\qquad
\Sigma=r^2+a^2\cos^2\theta,
\end{equation}
\begin{equation}
\omega=2Mra\sin^2\theta/(a^2\sin^2\theta-\Delta),
\end{equation}
($a$ is the Kerr rotation parameter, M is the mass),
one can write the transformed metric corresponding to the axidilaton
gravity in the same form, where the substitutions
\begin{equation}
\Delta_d \rightarrow \Delta, \qquad
\Sigma_d \rightarrow \Sigma,
\end{equation}
are to be done, where
\footnote{ The coordinate $r$ used here corresponds
to  $r_0$ in the definition of ref. \cite{gal}.}
\begin{equation}
\Delta_d=r(r+r_-) -2Mr + a^2 ,
\end{equation}
\begin{equation}
\Sigma_d=r(r+r_-)+ a^2\cos^2\theta,
\end{equation}
\begin{equation}
r_-=  Q^2/ M,
\end{equation}
$Q$ is the electric charge.
\par
It will be convenient for our analysis to represent the Kerr
solution to axidilaton gravity in the Kerr coordinates \cite{dks}.
We will do it in two steps by representation the charged Kerr solution
(the Kerr--Newman solution) at the first step
in the Boyer--Lindquist form \cite{mtw} in terms of parameters
$\Delta$ and $\Sigma$:
\begin{equation}
ds^2= - \frac{\Delta}{\Sigma}\left[t- a \sin^2\theta
d\varphi\right]^2 +\frac{\sin^2\theta}{\Sigma}\left[
(r^2 + a^2) d\varphi - a dt^2 \right]^2
\frac{\Sigma}{\Delta} {dr^2}+ \Sigma d\theta^2.
\end{equation}
The corresponding electromagnetic field is given by the vector potential
\footnote{The extra factor $2^{3/2}$ in the definition of the electric
charge has been introduced  to match definitions of refs. \cite{sen,gal} and
ref.\cite{mtw}.}
\begin{equation}
A =2^{3/2}Q \frac{r}{\Sigma} (dt - a \sin^2\theta d\varphi).
\end{equation}
\par
Next we rewrite the Kerr--Newman solution in the
Kerr coordinates by using the relations \cite{mtw}
\begin{equation}
 d \tilde V = dt - \frac{\Sigma + a^2\sin^2\theta}{\Delta} dr, \qquad
d\tilde\varphi = d\varphi + \frac{a}{\Delta} dr
\end{equation}
and express it again in  terms of the parameters $\Delta$ and $\Sigma$
\begin{equation}
ds^2= {\Sigma}(d\theta^2 + \sin^2\theta d\tilde\varphi^2) +
2 K (dr -a \sin^2\theta d\tilde\varphi) -(1-2H) K^2.
\end{equation}
Here
\begin{equation}
H=(\Sigma + a^2\sin^2\theta -\Delta)/\Sigma
\end{equation}
and K is a vector field tangent to the one of two principal null directions
of the Kerr solution
\begin{equation}
K= d{\tilde V} - a\sin^2 d\tilde \varphi.
\end{equation}
Electromagnetic field for the electric charged Kerr solution is
given by the vector potential
\begin{equation}
A = 2^{3/2}Q( r/\Sigma) K.
\end{equation}
After substituting $\Delta_d \rightarrow \Delta, \qquad
\Sigma_d \rightarrow \Sigma,$ the expressions (11)-(14) yield, according
to the Gal'tsov--Kechkin algorithm, the transformed Kerr solution to the
axidilaton gravity in the Kerr coordinates.
\footnote{Equivalence of these forms for
$\Delta_d$ and $\Sigma_d$ may also be verified  by direct calculations
by using the relations given in Appendix A.}
Now the gauge field is given by
\begin{equation}
A = 2^{3/2} Q e^{\Phi_0} (r/\Sigma_d) K,
\end{equation}
where $\Phi_0$ is the asymptotic value of the dilaton field.
The axion field $\Psi$ and the dilaton field $\Phi$ are joined in the complex
axidilaton field
\begin{equation}
\lambda= \Psi + i e^{-2 \Phi} = \lambda_0 + ir_- e^{-2\Phi_0}/
(r+ia \cos \theta),
\end{equation}
where
\begin{equation}
\lambda_0= \Psi_0 + i e^{-2 \Phi_0},
\end{equation}
is an asymptotic value of the axidilaton.
\par
This form allows us to use the Debney--Kerr--Schild (DKS) formalism
\cite{dks} to analyse the solution.
We represent metric of the transformed solution in tetrad form
\begin{equation}
ds_d^2 = 2 \widetilde e^3 \widetilde e^4 + 2 \widetilde e^1 \widetilde e^2
\end{equation}
and express it via the original DKS-tetrad $ e^a,  a=1,2,3,4, $
as a deformation of the Kerr solution by the dilaton factor
\begin{equation}
ds_d^2 = 2 e^3 e^4 + 2 e^1 e^2 e^{-2 (\Phi -\Phi_0)},
\end{equation}
where
\begin{equation}
 e^{-2 (\Phi- \Phi_0)}= \Sigma_d / \Sigma.
\end{equation}
Thus we have a new tetrad
\begin{equation}
\widetilde e^1 = e^{- (\Phi -\Phi_0)} e^1, \qquad \widetilde e^2 =
e^{- (\Phi -\Phi_0)} e^2,
\end{equation}
\begin{equation}
\widetilde e^3 =  e^3, \qquad \widetilde e^4 =  e^4 (with \quad substitution
H_d \rightarrow H),
\end{equation}
where the original DKS-tetrad is the following:
\footnote{The DKS-tetrad suffixes are
raised or lowered by performing the permutation $1,2,3,4
\rightarrow 2,1,4,3.$}
the tetrad null vectors $e^1$ and $e^2$ are complex conjugate
\begin{equation}
e^1 =2^{-1/2} Z^{-1} (d\theta + i \sin \theta d\tilde\varphi) =
(PZ)^{-1} d Y , \qquad e^2 = \bar e^1;
\end{equation}
 $e^3$ and $ e^4$ are real null vectors
\begin{equation}
e^3 = K , \qquad e^4 = dr + ia P^{-2} (\bar Y dY -Y d \bar Y ) + 2^{-1}
(H -1)e^3.
\end{equation}
{}From (12) we obtain the function $H_d$
\begin{equation}
H_d = 2Mr/ \Sigma_d.
\end{equation}
The functions $P$, $Z$ and $Y$ are
\begin{equation}
P = 2^{-1/2} (1+Y \bar Y), \qquad Z = (r + ia \cos \theta)^{-1},
\qquad Y = e^{i \tilde\varphi} \tan \theta /2.
\end{equation}
\par
\par
Now we would like  to get some  algebraic characteristics of the new
solution in comparison with the corresponding characteristics of the
original Kerr solution.
In the Kerr solution the vector $e^3=K$ is the multiple
Debever--Penrose vector tangent to a geodesic and shear free null congruence,
thus, the Kerr solution is algebraically special of type D in Petrov--Pirani
classification. The condition for
$e^3$ to be a Debever--Penrose null vector is expressed via the
component of Weyl's conformal curvature tensor \cite{dks}
\begin{equation}
C^{(5)} = 2 R_{4242} =0.
\end{equation}
The condition for $e^3$ to be a double Debever--Penrose vector
(or solution to be algebraically special ) is
\begin{equation}
C^{(4)} = R_{1242} + R_{3442} =0.
\end{equation}
The geodesic and shear-free condition for $e^3$ is
\begin{equation}
\Gamma_{424} = \Gamma_{422} =0.
\end{equation}
Checking these conditions for the axidilatonic Kerr
solution we obtain:
\footnote{In the Appendix B the expressions for the Ricci
rotation coefficients $ \widetilde {\Gamma^a_{bc}} $ are given via the known
values of the coefficients for the original Kerr solution $ \Gamma^a_{bc}$,
there are given also some necessary tetrad components of the curvature
tensor.}
\par
\medskip
i)
\begin{equation}
\widetilde{C^{(5)}} = 2 \widetilde{R_{4242}} =0,
\end{equation}
or $\tilde e^3$ is a Debever--Penrose null vector forming the principal null
congruence,
\par
\medskip
ii)
\begin{equation}
\widetilde{\Gamma_{424}} = \widetilde{\Gamma_{422}} =0,
\end{equation}
the principal null congruence of $\tilde e^3$ is geodesic and shear-free,
\par
\medskip
iii)
\begin{equation}
\widetilde{C^{(4)}} = \widetilde{R_{1242}} + \widetilde{R_{3442}} \neq 0,
\end{equation}
therefore, the new axidilatonic Kerr solution is not algebraically special,
 it is of type I in the Petrov--Pirani classification.
\par
\bigskip
\section{Limiting form of the axidilatonic Kerr solution near the
singular ring.}
\par
The Kerr singular ring is one of the remarkable peculiarities of the Kerr
solution.
It is a branch line of space on two sheets: "negative" and "positive" where
the fields change their signs and directions. There exist the Newton and
Coulomb analogues of the Kerr solution possessing the Kerr singular ring.
The corresponding Coulomb solution was obtained
by Appel still in 1887 (!) by a method of complex shift \cite{app}.
\par
A point-like charge $q$, placed on the complex Z-axis $(x_0,y_0,z_0)=
(0,0, ia)$
gives the real Appel potential
\begin{equation}
\phi_a = q/{\tilde r} + \bar q / \bar{\tilde r}.
\end{equation}
Here $\tilde r$ is in fact the Kerr complex radial coordinate
$Z^{-1}= r+ i a \cos\theta$. It may be expressed in the usual
rectangular Cartesian coordinates $x,y,z,t$  as
\begin{equation}
\tilde r \equiv Z^{-1}= [(x-x_0)^2 + (y-y_0)^2 + (z-z_0)^2]^{1/2}
= [x^2 + y^2 + (z-ia)^2]^{1/2}.
\end{equation}
It is not difficult to see that the Appel potential $\phi_a$ is
singular at the ring $z=0,\quad x^2+y^2=a^2$ or by $r=\cos \theta=0$.
\par
We would like to consider the axidilaton corrections to the original
Kerr field near the singular ring and will consider radius of the ring $a$
to be much larger than a distance $\delta$ from the singular line.
Thus, the parameter $\delta/a$ will be used as
a small parameter to get an approximate limiting form of metric near the
Kerr singularity.
\par
Formulas for the connection of the Cartesian and the Kerr angular
coordinates are the following
\begin{eqnarray}
x+iy &=& (r+ia) e^{i\tilde\varphi} \sin\theta,\\
z&=& r \cos\theta,\\
t&=& \tilde V - r .
\end{eqnarray}
By using these coordinates the Kerr metric may be expressed in the
Kerr--Schild form \cite{dks}
$g_{\mu\nu} = \eta_{\mu\nu} + 2h K_{\mu} K_{\nu},$
where $\eta$ is metric of the auxiliary Minkowski space.
The coordinates $r, \theta, \tilde \varphi$ cover the Minkowski space twice,
by positive and by negative values of $r$ with a branch line along the
singular ring $ r=\cos\theta=0 $; so coordinate $r$ will be
 two-valued near the Kerr singular filament.
\par
Near the point of singularity $ (x,y,z)=(a,0,0)$, in the orthogonal to the
filament 2-plane $y=0$ we introduce coordinates with origin on the filament
\begin{equation}
u= z,\quad v=x-a,
\end{equation}
and obtain from (34), keeping the leading term in $\delta/ a$
\footnote{Our approximation will be the most effective for the case of
a large $\mid a \mid$. The Kerr solution with $\mid a\mid  \gg  m$
has attracted a special attention because it displays some
relationships with the spinning elementary particles \cite{dks,bi,car,so}.
For example, the corresponding parameters of the electron will be
$a \approx 10^{22}, \quad m \approx 10^{-22}$ in units $\hbar=c=1$.
In this case all the fields concentrate very close to the singular filament.}
\begin{equation}
\tilde r = Z^{-1} \simeq a[2(v+iu)/a]^{1/2},
\end{equation}
\begin{equation}
d \tilde r  \simeq (dv+idu)/[2(v+iu)/a]^{1/2}.
\end{equation}
Function $Y$ in Cartesian coordinates may be extracted from eq.(5.72) of
\cite{dks}
\begin{equation}
Y= (z-ia - \tilde r)/ (x-iy),
\end{equation}
that yields
\begin{equation}
d Y \simeq (dz - d\tilde r)/a.
\end{equation}
By using the coordinate transformations (35)-(37) and the relations (39)-(42)
one finds the limiting form of tetrad (21)-(24) near the singular filament,
up to the leading terms in $\delta/a$
\begin{equation}
\widetilde e^1 = -  e^{-(\Phi -\Phi_0)} 2^{-1/2}(dv +i du) ,
\qquad \widetilde e^2 = -  e^{-(\Phi -\Phi_0)} 2^{-1/2}(dv -i du) ,
\end{equation}
\begin{equation}
\widetilde e^3 =  2^{-1/2} (dt - dy) , \qquad
\widetilde e^4 =  2^{-1/2} (dt + dy) + H_d 2^{-1/2}(dt - dy),
\end{equation}
where $ dy $ is directed along the singular filament. The functions
$\Sigma, \Sigma_d, \quad H_d$ and $ e^{-(\Phi -\Phi_0)}$ are given by
\begin{equation}
\Sigma \simeq 2a(v^2 + u^2)^{1/2} ,
\end{equation}
\begin{equation}
\Sigma_d \simeq 2a(v^2 + u^2)^{1/2} + a r_- \{[2(v+iu)/a]^{1/2}+
[2(v-iu)/a]^{1/2}\},
\end{equation}
\begin{equation}
H_d =2Mr/\Sigma_d,
\end{equation}
\begin{equation}
e^{-2(\Phi -\Phi_0)} = \Sigma_d/ \Sigma = 1 + r_-(Z + \bar Z)/2 .
\end{equation}
The limiting form of metric is
\begin{equation}
ds_d^2 = e^{-2(\Phi -\Phi_0)}(dv^2 + du^2) + dy^2 - dt^2  +
(2Mr/\Sigma_d) (dy -dt)^2.
\end{equation}
The gauge field is given by the vector potential
\begin{equation}
A =  2 Q (r/\Sigma_d)(dt - dy).
\end{equation}
By introducing the electric charge per unit length of the Kerr ring $ q=
2^{(3/2)}Q/(2 \pi a)$ and a two-dimensional (two-valued) Green's function
$G_a^{(2)}$ in the $(u,v)$ complex plane near the Kerr singularity
\begin{equation}
 G_a^{(2)}= 2 \pi ar/\Sigma \simeq  \pi \{[\frac{a}{2(u+iv)}]^{1/2} +
  [\frac {a}{2(u-iv)}]^{1/2}\},
\end{equation}
the dilaton factor may be represented as
\begin{equation}
e^{-2(\Phi -\Phi_0)} = 1 + N G_a^{(2)} ,
\end{equation}
where
\begin{equation}
N =r_- /2\pi a.
\end{equation}
Then the rescaled $\sigma$-model metric
$ ds_{str}^2 =  e^{2(\Phi -\Phi_0)}ds_d^2, $
used in string theory may be written in the form
\begin{equation}
ds_{str}^2 = (dv^2 + du^2) +
\frac{1}{1+ NG_a^{(2)}}
(dy^2 - dt^2) + \frac{2M G_a^{(2)}}{2\pi a(1+ NG_a^{(2)})^2} (dy -dt)^2.
\end{equation}
This metric coincides with the form of metric obtained by Sen
for the field around a fundamental heterotic string \cite{shet,sen}.
\footnote{Sen has constructed this solution by the method for
generating new solutions from the fundamental string solution
of ref.\cite{dab}.} It allows to identify the Kerr singular ring
in axidilaton gravity  with a heterotic string.
\par
However, there is one peculiarity of this solution that two-dimensional
Green's function $G_a^{(2)}$
differs from the Green function of the Sen solution corresponding to
the simple line source. This difference is very natural and it is connected
with the twovaluedness of the fields near the Kerr singularity and with
the known twofoldidness of the Kerr space.
\par
This twovaluedness was an object of the special consideration in the old
problem of the source of the Kerr solution \cite{so}. One of the  traditional
solutions of this problem is  cutting off the negative
sheet of the Kerr space by introducing a disk-like source spanned by
the Kerr singular ring. The analysis shows \cite{so} that this disk has to
be in a rigid relativistic rotation and consists of an exotic material
with superconducting properties.
Thus, the  Kerr heterotic string is to be placed at the board of the
superconducting disk that is in agreement with the
superconducting nature of the heterotic strings mentioned before in refs.
\cite{shet,wit}.
Some earlier presumptions concerning the Kerr singular ring to be a string
may be found in refs.\cite{bi}.
\par
Further, it would be interesting
to consider electromagnetic and axidilatonic excitations of the Kerr string
in the form of traveling waves \cite{gar},
\footnote{Similar model for the Kerr
solution in Einstein's gravity  was suggested in refs. \cite{geon}.}
and the case of massive dilaton.
\par
There is one more string-like structure in the Kerr geometry which is
connected with the above representation of the Kerr source as an object
propagating along a complex world line and based on the fact that the
complex world line is really a world sheet \cite{cstr}. The physical role
of these strings and their interaction are still unclear.
\par
\bigskip
{\bf Note added}: After this paper was written I was informed that
the Petrov-Pirani type of the Kerr solution in axidilaton gravity was
also determined  by Gal'tsov and Lunin (unpublished).
\par
In conclusion I would like to thank D. Gal'tsov for useful discussions.
\par
\bigskip
{\bf Appendix A}
\par
To match the notations of refs. \cite{sen} and \cite{gal} we will add
 subscripts $s$ for the Sen parameters and $g$ for the Gal'tsov--Kechkin
parameters. Then we have
\begin{eqnarray}
Q_s&=&Q_g=Q, \qquad M=M_s= M_g= m_s \cosh^2\frac{\alpha_s}{2},\\
q&=&2\sqrt{2}Q\\
m_s&=& M-\frac{r_-}{2},\\
r_-&=&Q^2/M = 2 m_s \sinh^2\frac{\alpha_s}{2} =2(M_s-m_s)
\end{eqnarray}
The following relations are  useful when  deriving the transformed
solution in the Kerr coordinates
\begin{equation}
(\Sigma_d +a^2\sin^2\theta)^2 - \Delta_d a^2\sin^2\theta =
(\Sigma_d +a^2\sin^2\theta)\Sigma_d + 2Mr a^2 \sin^2\theta,
\end{equation}
\begin{equation}
dt - a\sin^2\theta d\varphi= K - \frac{\Sigma_d}{\Delta_d} dr,
\end{equation}
where the vector K is given by
\begin{equation}
K= d \tilde V - a\sin^2 d\tilde \varphi
\end{equation}
 and points in the principal
null direction.
In the Kerr coordinates
\begin{equation}
\tilde V = dt + dr,
\end{equation}
if the principal null congruence directed "inside".
\par
In the expression for vector potential (15) the term
$\frac{q}{\Sigma_d}\frac{\Sigma_d}{\Delta_d} dr$ is omitted
since it is full differential.
\par
\bigskip
{ \bf Appendix B}
\par
We use  a freedom of  tetrad transformations (Eqs.(2.21) of ref. \cite{dks})
to adopt the tetrad (23),(24) to the DKS-form of sec.3 of ref. \cite{dks}.
\begin{equation}
e'^1 = e^1, \quad e'^2=e^2,\quad e'^3 =P e^3, \quad e'^4 = P^{-1}e^4,
\quad Z'= PZ.
\end{equation}
Dropping primes, we calculate the Ricci rotation coefficients to the
axidilaton solution
expressed via the coefficients of the original Kerr solution $\Gamma_{abc}$.
For example, we extract $ \widetilde {\Gamma_{2 b c}}$ from the relations
\begin{equation}
\widetilde e^1 = e^{-(\Phi -\Phi_0)} e^1, \qquad d \widetilde e^1=
\widetilde
{\Gamma^1_{bc}} \tilde e^b \wedge \tilde e^c = e^{-(\Phi -\Phi_0)}
(d e^1 -d \Phi\wedge e^1) ,
\end{equation}
where
$ d e^1= \Gamma_{2 [ b c ]} e^b \wedge e^c.$
\par
The result is given by
\par
\bigskip
$\widetilde{\Gamma_{121}} = - e^{(\Phi -\Phi_0)} {\bar Z}
( {\bar Z} ^{-1} ),_{1} +
e^{(\Phi -\Phi_0)} \Phi,_{\tilde{1}},$
\par
$\widetilde{\Gamma_{122}} =  e^{(\Phi -\Phi_0)} Z( Z^{-1} ),_2
- e^{(\Phi -\Phi_0)} \Phi,_{\tilde2},$
\par
$\widetilde{\Gamma_{123}} =  e^{(\Phi -\Phi_0)}  {\Gamma_{123}}
 + e^{(\Phi -\Phi_0)} (1- e^{(\Phi -\Phi_0)}) \left[  (\Gamma_{312}  -
\Gamma_{321})/2 + (H-H_d) (Z - \bar Z )/4 \right] ,$
\par
$\widetilde{\Gamma_{124}} =  e^{(\Phi -\Phi_0)} (1- e^{(\Phi -\Phi_0)})
(Z - \bar Z )/2,$
\bigskip
\par
$\widetilde{\Gamma_{311}} = \widetilde{\Gamma_{314}} = 0,$
\par
$\widetilde{\Gamma_{312}} =  e^{(\Phi -\Phi_0)} (1+ e^{(\Phi -\Phi_0)} )
\Gamma_{312}/2  -
e^{(\Phi -\Phi_0)} (1- e^{(\Phi -\Phi_0)}) \Gamma_{321}/2
+(H-H_d)(e^{(\Phi -\Phi_0)} +1)(\bar Z - Z),$
\par
$\widetilde{\Gamma_{313}} = -(H-H_d),_{\tilde1}+ e^{(\Phi -\Phi_0)}
[\Gamma_{313} +(H-H_d) Z(Z^{-1}),_{2} ] ,$
\par
\bigskip
$\widetilde{\Gamma_{341}} =  - e^{(\Phi -\Phi_0)} \bar Z
({\bar Z}^{-1}),_{1}, $
\par
$\widetilde{\Gamma_{342}} =  - e^{(\Phi -\Phi_0)}  Z(Z^{-1}),_{2}, $
\par
$\widetilde{\Gamma_{343}} = \Gamma_{343} +(H-H_d),_{\tilde4}/2,$
\par
$\widetilde{\Gamma_{344}} = 0,$
\par
\bigskip
$\widetilde{\Gamma_{421}} = -  Z e^{(\Phi -\Phi_0)} (1+
e^{(\Phi -\Phi_0)} )/2 -
\bar Z e^{(\Phi -\Phi_0)} (1- e^{(\Phi -\Phi_0)} )/2, $
\par
$\widetilde{\Gamma_{422}} = \widetilde{\Gamma_{423}} =
\widetilde{\Gamma_{424}} = 0.$
\par
Directional derivatives along the tetrad vectors are
$,_a = ,_{\mu} e^{\mu}_a$ and
 $,_{\tilde a} = ,_{\mu} {\tilde e}^{\mu}_a$.
\bigskip
The  curvature tensor is defined by the Cartan formula
\begin{equation}
 {\cal R}^a_b = R^a_{bcd} e^c \wedge e^d
 = d \Gamma^a_b + \Gamma^a_m\wedge \Gamma^m_b.
\end{equation}
Some tetrad components of the curvature tensor for the axidilatonic Kerr
solution are
\begin{equation}
\widetilde {R_{4242}} =
\widetilde{ R_{4234}} =
\widetilde {R_{4223}} = 0,
\end{equation}
\begin{equation}
\widetilde {R_{4214}} = (\chi^2 - \chi,_{\tilde 4})/2,
\qquad \widetilde {R_{4212}} = \chi ( \widetilde{\Gamma_{212}} -
2 e^{(\Phi -\Phi_0)} \Phi,_{\tilde 2}) - 2 \chi,_{\tilde 2},
\end{equation}
where
\begin{equation}
\chi =-(1/2) e^{(\Phi -\Phi_0)}[Z(1+ e^{(\Phi -\Phi_0)}) + \bar Z
(1- e^{(\Phi -\Phi_0)})],
\end{equation}
and
$ Z=(r+ia\cos \theta)/P $.
\par

\end{document}